\documentclass[conference]{IEEEtran}
\IEEEoverridecommandlockouts
\usepackage{cite}
\usepackage{fontawesome}
\usepackage{float}
\usepackage{graphicx}
\usepackage{subfigure}
\usepackage{natbib}
\usepackage{amsmath,amssymb,amsfonts}
\usepackage{algorithmic}
\usepackage{graphicx}
\usepackage[table]{xcolor}
\usepackage[dvipsnames]{xcolor}
\usepackage{textcomp}
\usepackage{lipsum}
\usepackage{mwe}
\usepackage{xcolor}
\usepackage{subfigure}
\usepackage{subcaption}  
\usepackage[colorlinks=true,linkcolor=blue,citecolor=blue,urlcolor=blue]{hyperref} 
\usepackage{multirow}
\usepackage{float}
\def\BibTeX{{\rm B\kern-.05em{\sc i\kern-.025em b}\kern-.08em
    T\kern-.1667em\lower.7ex\hbox{E}\kern-.125emX}}

\begin{document}

\title{A Benchmark Dataset And LLMs Comparison For NFR Classification With Explainable AI\\
}

\author{
\begin{minipage}[t]{0.35\textwidth}
\centering
Esrat Ebtida Sakib\\
\textit{Dept. of CSE}\\
\textit{Islamic University of Technology}\\
Gazipur, Bangladesh\\
esrat@iut-dhaka.edu
\end{minipage}
\begin{minipage}[t]{0.35\textwidth}
\centering
MD Ahnaf Akib\\
\textit{Dept. of CSE}\\
\textit{Islamic University of Technology}\\
Gazipur, Bangladesh\\
akib@iut-dhaka.edu
\end{minipage}\hfill
\begin{minipage}[t]{0.35\textwidth}
\centering
Md Muktadir Mazumder\\
\textit{Dept. of CSE}\\
\textit{Islamic University of Technology}\\
Gazipur, Bangladesh\\
muktadir@iut-dhaka.edu
\end{minipage}\hfill

\\[1em] 

\makebox[\textwidth][c]{%
\begin{minipage}[t]{0.35\textwidth}
\centering
Maliha Noushin Raida\\
\textit{Dept. of CSE}\\
\textit{Islamic University of Technology}\\
Gazipur, Bangladesh\\
malihanoushin@iut-dhaka.edu
\end{minipage}\hspace{2em} 
\begin{minipage}[t]{0.35\textwidth}
\centering
Md. Mohsinul Kabir\\
\textit{Dept. of CSE}\\
\textit{Islamic University of Technology}\\
Gazipur, Bangladesh\\
mohsinulkabir@iut-dhaka.edu
\end{minipage}%
}
}

\maketitle

\begin{abstract}

Non-Functional Requirements (NFRs) play a critical role in determining the overall quality and user satisfaction of software systems. Accurately identifying and classifying NFRs is essential to ensure that software meets performance, usability, and reliability expectations. However, manual identification of NFRs from documentation is time-consuming and prone to errors, necessitating automated solutions.

Before implementing any automated solution, a robust and comprehensive dataset is essential. To build such a dataset, we collected NFRs from various Project Charters and Open Source Software Documentation. This enhanced the technical depth and usability of an already-existing NFR dataset. We categorized NFRs into sub-classes and identified needs using widely used Large Language Models to facilitate automation. After classifying the NFRs, we compared the classification results of the selected LLMs - \textit{RoBERTa, CodeBERT, Gemma-2, Phi-3, Mistral-8B, and Llama-3.1-8B} models using various evaluation metrics, including precision, recall, F1-score, and lime scores. Among these models, Gemma-2 achieved the best results with a precision of 0.87, recall of 0.89, and F1-score of 0.88, alongside a lime hit score of 78 out of 80. Phi-3 closely followed with a precision of 0.85, recall of 0.87, F1-score of 0.86, and the highest lime hit score of 79.

By improving the contextual foundation, this integration enhanced the model's comprehension of technical aspects and user requirements.

\end{abstract}

\section{Introduction}
The quality of the software development process primarily depends on the software requirements. These requirements represent what users expect from the software and are usually provided by stakeholders or software owners. They are documented in project charters, which are formal documents that outline a project's objectives, scope, stakeholders, and deliverables.  Requirement elicitation identifies user needs and is documented through Software Requirement Specifications (SRS) and Project Charters. These specify system functions, behaviors, and constraints like security and performance~\cite{chung2012non}. Requirements are mainly classified into Functional Requirements (FRs) and Non-functional Requirements (NFRs). FRs define system operations~\cite{malan2001functional}, while NFRs describe quality attributes like performance and security~\cite{ieee1983ieee}. 
With the proliferation of mobile and internet experiences, modern software applications must adapt to various factors, such as the user's location, intent, and the time of day. This evolution means that applications must consider non-functional aspects such as performance, usability, and security, in addition to their functional requirements. By addressing these requirements early in the development process, developers can ensure the creation of high-quality software that meets both user expectations and project goals \cite{lu2017automatic}.

 \subsection{Motivation}
\label{sec:motivation}

Since non-functional requirements (NFRs) are often overlooked, it is crucial to identify them early in the software development process. Traditionally, this task has involved manually classifying requirements, a process that can be time-consuming, inaccurate, and dependent on domain-specific knowledge—often scarce and expensive. For example, expertise in security is required to identify potential vulnerabilities in critical systems, and its absence can lead to missed flaws that pose significant risks.

However, the automatic identification of non-functional requirements offers significant advantages to the software community. In this context, project charters play a pivotal role. These documents typically capture key quality attributes and expectations early on, providing valuable insight into non-functional requirements. Despite the clear value of project charters, there is a gap in the availability of comprehensive and labeled datasets from these documents, making it an underexplored yet promising area for research in software engineering.

 \subsection{Our Contributions}
 \label{sec:contributions}



 


The key contributions of this study are outlined as follows:

\begin{itemize}
    \item We extended the benchmark dataset named PROMISE\_exp dataset \cite{lima2019software}, which addresses software requirements, by categorizing these requirements into more than six Non-functional Classes and incorporating new types of Non-functional Requirements derived from project charters and Open Source GitHub Repositories, expanding the range of NFR categories.
    
    \item We provided definitions and examples of each Non-functional Requirement (NFR) category to ensure clarity and better understanding. NFRs were classified based on these definitions.
    
    \item We leveraged Large Language Models (LLMs) such as RoBERTa, CodeBERT, Gemma-2, Phi-3, Mistral-8B, and Llama-3.1-8B to classify these NFRs accurately and distinguish between various NFR types.
    
    
    \item We compared the performance of the selected LLMs, utilizing Explainable AI techniques to gain insights into model decisions and ensure transparency in the classification process.
\end{itemize}

\section{Literature Review}
\label{sec:lit}


\subsection{Deep Learning Approaches}
Classifying non-functional requirements has been greatly improved by recent advances in deep learning. In order to overcome issues like ambiguity and inconsistency, Gramajo et al. (2021) \cite{gramajo2021recurrent} presented a novel method that uses Recurrent Neural Networks (RNNs) to automatically evaluate the quality of software requirements written in natural language. Adding to this, Rahman et al. (2019) \cite{rahman2019classifying} investigated the application of RNN variations for high-quality software development, proving the efficacy of deep learning techniques in classifying NFRs and obtaining remarkable accuracy. Furthermore, Hey et al. (2020) \cite{hey2020norbert} proposed NoRBERT, a transfer learning model that improves generalization across projects by fine-tuning the BERT framework for high F1-scores in classifying functional and non-functional requirements. 

\subsection{Machine Learning Approaches}
Several works in the field of machine learning have concentrated on improving the categorization of non-functional requirements (NFRs). Through high accuracy, the transfer learning model proposed by Khan \emph{et al.} (2023) \cite{khan2023non} helps to increase software development efficiency by identifying and classifying NFRs. An empirical study using a variety of machine learning methods for categorizing NFRs was carried out by Haque \emph{et al.} (2019) \cite{haque2019non}, who provided insightful information on feature extraction and how it affects classification performance. In addition, in their exploration of several machine learning methods, Rahman \emph{et al.} (2023) \cite{rahman2023non} highlighted the efficacy of various classifiers and vectorization strategies in NFR classification. Kurtanovic \emph{et al.} (2017) \cite{kurtanovic2017automatically} used supervised learning approaches to automatically classify software requirements into functional and non-functional categories. The efficiency of linear classifiers and deep learning models is demonstrated by their work and the study of Tóth \emph{et al.} (2019) \cite{toth2019comparative}, which examined the performance of different classifiers in labeling NFRs. This is especially true when there is an adequate amount of training data available.

\subsection{Explainable AI}
Our study is closely related to the paper ``Bridging Interpretability and Robustness Using LIME-Guided Model Refinement''\cite{lime_guided_refinement} since it emphasizes how LIME can assess feature importance and improve models by detecting spurious dependencies. In a similar vein, our study uses LIME to evaluate NFR classification models and make sure they emphasize significant aspects.

The article titled ``Why Model Why'' \cite{why_model_why}, since it examines LIME's function in assessing model interpretability, ``Assessing the Strengths and Limitations of LIME'' directly relates to our study. Like our method, it demonstrates how LIME can uncover influential elements to explain machine learning predictions. This promotes improving transparency and dependability by verifying model decisions against anticipated patterns.

\subsection{Data Sets for NFR Classification}
There are 625 labeled natural language requirements in the open-source tera-PROMISE repository, which are divided into 255 functional and 370 non-functional requirements \cite{xuapplying}. To create this information, needs from fifteen distinct projects were gathered. Non-functional requirement categories that are represented include usability, fault tolerance, availability, legal considerations, look and feel, maintainability, operability, performance, scalability, security, and portability.

The tera-PROMISE non-functional requirements (NFR) dataset was deemed too small, so Lima \emph{et al.} (2019) \cite{lima2019software} created the expanded PROMISE\_exp dataset, which includes 525 non-functional requirements and 444 functional requirements. 
By improving the classification of non-functional requirements using cutting-edge machine learning and deep learning approaches, these studies collectively advance the field of requirements engineering. 

\section{Methodology}
\label{sec:methodology}


\begin{figure*}[htbp]
    \centering
    \includegraphics[width=1.0\textwidth]{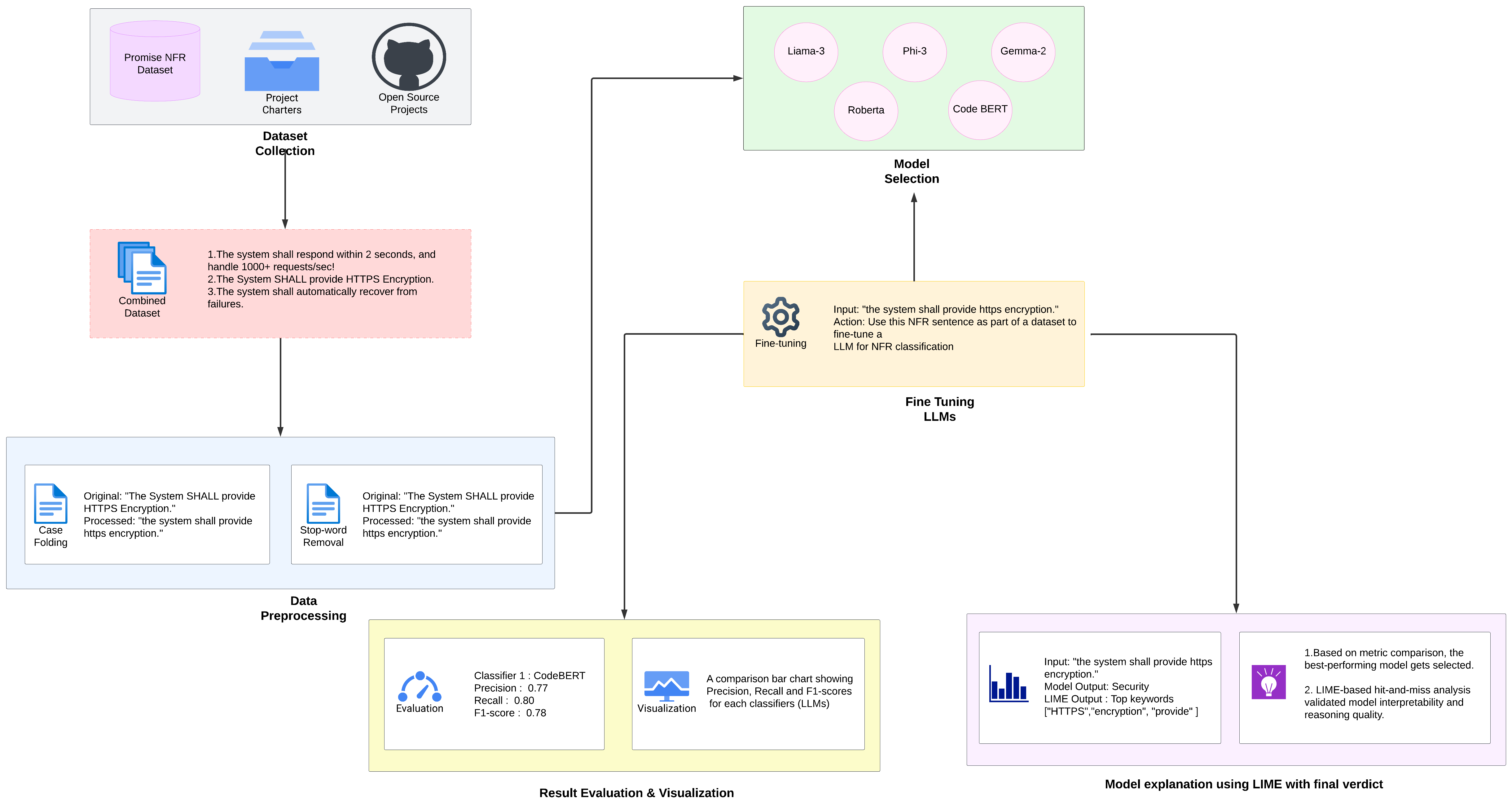}
    \caption{Proposed Methodology}
    \label{proposed_methodology}
\end{figure*}

\subsection{Dataset Overview and Pre-processing}\label{subsec:dataset}


The extended PROMISE\_exp dataset \cite{lima2019software}, which had 444 functional and 525 non-functional requirements, was used in our investigation. Following a comprehensive process of human analysis, extraction, and classification, the authors used the Google search engine to collect Software Requirements Specification (SRS) papers. Expert consensus was then used to validate the resultant requirement categories. The following table~\ref{tab:combined_requirements} shows the distribution of needs for each category in the extended dataset.

\begin{table}[htbp]
\centering
\renewcommand{\arraystretch}{1.2}
\setlength{\tabcolsep}{8pt}
\begin{tabular}{|c|c|c|}
\hline
\hline
\multirow{2}{*}{\textbf{Requirement Type}} & \multicolumn{2}{c|}{\textbf{Number of Requirements}} \\ 
\cline{2-3}
& \textbf{PROMISE NFR} & \textbf{Extended NFR} \\ 
\hline
Availability (A)          & 31                  & 44                    \\ 
\hline
Legal (L)                 & 15                  & 24                    \\ 
\hline
Look-and-feel (LF)        & 49                  & 49                    \\ 
\hline
Maintainability (MN)      & 24                  & 138                   \\ 
\hline
Operability (O)           & 77                  & 154                   \\ 
\hline
Performance (PE)          & 67                  & 118                   \\ 
\hline
Scalability (SC)          & 22                  & 25                    \\ 
\hline
Security (SE)             & 125                 & 358                   \\ 
\hline
Usability (US)            & 85                  & 165                   \\ 
\hline
Fault Tolerance (FT)      & 18                  & 19                    \\ 
\hline
Portability (PO)          & 12                  & 14                    \\ 
\hline
\textcolor{Tan}{Interoperability (I)}      & -                   & 2                     \\ 
\hline
\textcolor{Tan}{Compliance (C)}            & -                   & 5                     \\ 
\hline
\textcolor{Tan}{Compatibility (CM)}        & -                   & 2                     \\ 
\hline
\textcolor{Tan}{Reliability (R)}           & -                   & 4                     \\ 
\hline
\hline
\end{tabular}
\caption{Comparison of PROMISE NFR Dataset and Combined NFR Dataset.}
\label{tab:combined_requirements}
\end{table}

For our research on Non-Functional Requirements (NFRs) classification, we chose project charters from \textbf{Battery Low Interactive}\footnote{Battery Low Interactive official website: \url{https://batterylowinteractive.com}} and \textbf{Perceptron}\footnote{Perceptron official website: \url{https://www.perceptron.site}} — two renowned software development companies in Bangladesh. Whereas \textit{Perceptron} focuses on developing online and mobile apps, \textit{Battery Low Interactive} primarily creates games.

We also gathered NFRs from open-source projects on GitHub, including \textbf{Refine}\footnote{Refine repository: \url{https://github.com/refinedev/refine}}, \textbf{COVID Tracker}\footnote{COVID Tracker repository: \url{https://github.com/arafaysaleem/covid_tracker}}, \textbf{OSMCha Frontend}\footnote{OSMCha Frontend repository: \url{https://github.com/OSMCha/osmcha-frontend}}, and \textbf{TODO Group Governance}\footnote{TODO Group Governance repository: \url{https://github.com/todogroup/governance}}.

This approach emphasizes the use of real-world data to develop robust NFR classifiers, enhancing the realism and applicability of our findings. Definitions and examples for each NFR category are provided in our GitHub\footnote{\href{https://github.com/acesuu/A-Benchmark-Dataset-And-LLMs-Comparison-For-NFR-Classification-With-Explainable-AI/tree/main/Definition}{GitHub Repository – NFR Definitions and examples}}.The new NFRs in table~\ref{tab:combined_requirements} that are founded from the project charters and open source projects are \textit{{Compatibility (CM)}}, \textit{{Interoperability (I)}}, \textit{{Compliance (C)}}, \textit{{Reliability (R)}} and \textit{{Localization (L)}}. 

The general research technique employed in this paper is depicted in figure~\ref{proposed_methodology}. The procedure includes pre-processing the dataset, putting the data into training embedding, choosing a model, training the model, and assessing the model, among other steps. The first step involves cleaning the dataset. We used specific prompts to help classify non-functional requirements (NFRs), as shown in Table~\ref{tab:nfr_prompts}.

\begin{table*}[h]
\centering
\small
\caption{Input prompts used for different NFR categories}
\begin{tabular}{|p{16cm}|}
\hline
\multicolumn{1}{|c|}{\textbf{Prompt}} \\
\hline
Below is an instruction that describes a task, paired with an input that provides further context. Write a response that appropriately completes the request. \\
\textbf{Instruction:} Given the sentence, classify the Non-Functional Requirement (NFR) into one of the following categories: Performance, Scalability, Compatibility, Fault Tolerance, Legal, Maintainability, Operability, Portability, Security, Usability, Reliability, Interoperability, Compliance, Look-and-Feel, Availability.  \\
\textbf{Input:} The website shall continue to operate if the payment gateway goes down. \\
\hline

\hline
\end{tabular}

\label{tab:nfr_prompts}
\end{table*}

\subsection{Model Selection and Training}

CodeBERT \cite{codeberta}, fine-tuned for NFR categorization, extracts semantic relationships from code, enhancing NFR identification. Gemma-2 \cite{team2024gemma}, optimized for lightweight performance, excels with a smaller dataset, making it ideal for resource-constrained environments. Phi-3 \cite{abdin2024phi}, trained on 3.3 trillion tokens, is efficient for code summarization tasks, with 4-bit quantization for mobile devices. RoBERTa \cite{liu2019roberta}, improves on BERT by using larger data and better pretraining for robust NLP performance. Mistral \cite{jiang2023mistral}, with advanced attention mechanisms, is designed for long sequences and handheld deployment. Llama-3 \cite{touvron2023llama}, pretrained on 15 trillion tokens, excels in high-performance, resource-efficient tasks, using reinforcement learning from human feedback (RLHF) for fine-tuning.

By fine-tuning, these models learn to detect subtle cues and associate code properties with specific NFR categories.
To increase classification accuracy, we began utilizing a larger dataset to train our CodeBERT, RoBERTa, Gemma-2, Phi-3, Mistral-8B and Llama-3.1-8B classifiers. Each classifier has its hyper-parameters, such as epochs, batch sizes, and learning rates, carefully calibrated.
Each trained classifier produces an output for a requirement regardless of whether it fits into a category. The classifiers enable efficient classification and analysis of the given needs by producing outputs that indicate the classified NFR categories, such as performance, security, portability, or others, when given a query non-functional requirement (NFR) statement as input.

\subsection{Step 04: Classifying requirements}
Each trained classifier produces an output for a requirement regardless of whether it fits into a category. The classifiers enable efficient classification and analysis of the given needs by producing outputs that indicate the classified NFR categories, such as performance, security, portability, or others, when given a query non-functional requirement (NFR) statement as input.




\subsection{Step 05: Evaluation Metrics}

We used the following evaluation metrics to assess our classifiers numerically:
\begin{enumerate} 
    \item \textbf{Precision:}
\textit{Precision} evaluates the accuracy with which a classification model forecasts the favorable results. It is the ratio of real positives, or accurately predicted positives, to all forecasted positives, or false positives, put together. It answers the question,``How many of the occurrences that the model classified as positive are actually positive'' to put it simply.

Mathematically, Precision is defined as:
\begin{equation*}
\text{Precision} = \frac{\text{True Positives}}{\text{True Positives} + \text{False Positives}}
\label{Precision}
\end{equation*}

\item \textbf{Recall:}
\textit{Recall} assesses the model's ability to identify each pertinent event; it is also known as the True Positive Rate or Sensitivity. It is defined as the ratio of real positives to actual positives, including false negatives and true positives. ``How many actual positives did the model correctly identify?'' is a question that recall answers.

Mathematically, Recall is defined as:
\begin{equation*}
\text{Recall} = \frac{\text{True Positives}}{\text{True Positives} + \text{False Negatives}}
\label{Recall}
\end{equation*}

 \item\textbf{F1 Score:}
    Precision and recall are balanced by the \textit{F1 Score}, which is the harmonic mean of the two measures. It is particularly useful when the distribution of classes is unbalanced. The F1 Score is at its highest at 1 (perfect recall and precision) and lowest at 0.

Mathematically, F1 Score is defined as:
\begin{equation*}
F1 \text{ Score} = 2 \times \frac{\text{Precision} \times \text{Recall}}{\text{Precision} + \text{Recall}}
\label{ F1-Score}
\end{equation*}

\item\textbf{Lime Score Calculation and Visualization}
To comprehend how NFR classification models make decisions, we used LIME (Local Interpretable Model-agnostic Explanations). Features (such words) are given interpretability ratings by LIME, which show how much of an impact they have on predictions. By guaranteeing that predictions matched the right feature (word), this method verified the model's emphasis on essential keywords and improved the validity of our assessment.

\end{enumerate}

\subsection{Evaluation and Results Analysis}
We evaluated our classifiers using Precision, Recall, F1-Score and LIME. Precision measures the accuracy of positive predictions, while Recall evaluates the model's ability to identify relevant events. The F1-Score balances Precision and Recall, useful for imbalanced classes. LIME provided insights into feature importance and validated the model’s focus on key terms. We carried out a comprehensive analysis to compare model performance and included graphics to show the findings. LIME scores were also utilized to demonstrate how successfully each model identified NFR classes by concentrating on key terms.

\section{Experimental Setup}
\label{setup}

\subsection{Environment}
We used the free edition of Google Colab, which provided access to NVIDIA Tesla K80 GPUs, for model training. While not as powerful as specialized hardware, Colab facilitated the training and evaluation of CodeBERT, RoBERTa, Gemma-2, Phi-3, Mistral-8B, and Llama-3.1-8B by leveraging GPU acceleration. 

\subsection{Train-Test Split}
For both training and evaluation, we utilized the 80-20 train-test split technique. This means that 80\% of the dataset was used to train the transfer learning model, with the remaining 20\% being saved for performance assessment.

\subsection{Hyper Parameters and Fine-Tuning}
A per-device batch size of 2 was combined with gradient accumulation steps of 4, resulting in an effective batch size of 8. The maximum sequence length was determined by \texttt{max\_seq\_length} to handle long text inputs effectively. A linear learning rate scheduler was employed with an initial learning rate of \(2 \times 10^{-4}\), along with 5 warmup steps to stabilize training dynamics. Fine-tuning was performed for 60 steps using the AdamW optimizer with 8-bit precision, while weight decay of 0.01 was applied for regularization. Mixed precision training (FP16 or BF16) was utilized based on hardware support to improve computational efficiency. For fine-tuning, the SFTTrainer framework was employed to adapt the pre-trained model to the task-specific dataset. Dataset preprocessing leveraged parallelism (\texttt{dataset\_num\_proc = 2}), and sequence packing was disabled to support variable-length sequences.

\subsection{Lime score calculation and visualization}
We use LIME scores to assess whether models rely on the correct words when classifying sentences into NFR classes. We manually label keywords in a sample of 80 sentences, then compare them with the model’s highlighted words to identify discrepancies. What we want to convey by using lime score is how much each model is focussing on the right features (in our case - words) to take decision on which NFR class a sentence may fall.
We manually label keywords in a sample of 80 sentences, then run our LIME-based extractor to identify words that differ from our initial assumptions.

\section{Result}
\label{sec:result}

\subsection{Evaluation Metrics Result}
Running the selected classifiers on batch size = 08 and epoch = 03 we have found the following outcome in table~\ref{tab:eval_comparison}
\begin{table}[H]
\centering
\begin{tabular}{|c|c|c|c|}
\hline
\textbf{Classifiers} & \textbf{Precision} & \textbf{Recall} & \textbf{F1-score}\\
\hline
\hline
CodeBERT & 0.77 & 0.80 & 0.78 \\
RoBERTa & 0.29 & 0.32 & 0.30\\
Gemma-2 & \textbf{0.87}  &  \textbf{0.89} &  \textbf{0.88} \\
Phi-3 & 0.85  & 0.87 & 0.86 \\
Mistral-8B & 0.85 &   0.85 &   0.84 \\
Llama-3.1-8B &  0.83  &  0.85  & 0.84  \\
\hline
\end{tabular}
\caption{Evaluation Metrics Comparison}
\label{tab:eval_comparison}
\end{table}

By looking at the performance scores of the models it is evident that Gemma-2 outperforms RoBERTa, CoeBERT, Phi-3, Mistral-8B, and Llama-3.1-8B across all metrics, achieving the highest precision, recall, and F1-score. In contrast, RoBERTa demonstrates the lowest performance among all classifiers, while CoeBERT marginally outperforms RoBERTa.


\subsection{Evaluation of the lime score}
 LIME produced a list of words and phrases that affected the model's choice for each sentence, giving each one a score determined by how important it was. When the model's emphasis successfully matched the human-identified terms, a ``hit'' was noted. On the other hand, a ``miss'' happened when the model prioritized irrelevant or unrelated features more than others, indicating places where its logic and interpretability were weak. The models' capacity to give priority to significant patterns over noise was better understood as a result of this work.\\

\begin{table}[H]
\centering
\begin{tabular}{|c|c|c|c|}
\hline
\textbf{Classifiers} & \textbf{Hit} & \textbf{Miss} \\
\hline
CodeBERT & 77 & 3  \\
RoBERTa & 72 & 8 \\
Phi-3 & \textbf{79}  & \textbf{1}  \\
Mistral-8B & 77 &   3  \\
Llama-3.1-8B &  78  &  2    \\
\hline
\end{tabular}
\caption{LIME score comparison}
\label{tab:lime_comparison}
\end{table}

\begin{figure}
\begin{minipage}[t]{0.5\columnwidth}
  \includegraphics[width=\linewidth]{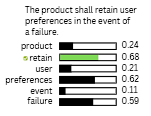}
  \caption{An example of hit}
   \label{fig:hit-image}
\end{minipage}\hfill 
\begin{minipage}[t]{0.5\columnwidth}
  \includegraphics[width=\linewidth]{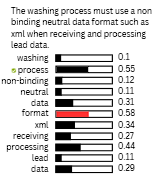}
  \caption{An example of miss}
  \label{fig:miss-image}
\end{minipage}
\end{figure}

To clear up the \textit{hit \& miss} analogy, in figure \ref{fig:hit-image} it can be seen that the sentence ``The product must retain user preferences in the event of failure'' ignoring the predefined stop words, the lime analysis produced the lime score presented in the figure. It can be observed that the word ``retain'' has the highest lime value that is 0.68, meaning, when the Phi-3 model came to the decision that this falls under the ``Availability'' NFR class, it focused on the feature/word ``retain'' the most, which is correct based on expert opinion. In Figure \ref{fig:miss-image}, an example of the \textit{miss} analogy is shown. The sentence ``The washing process must use a non-binding neutral data format such as XML when receiving and processing lead data'' should be classified under ``Interoperability'', which Phi-3 correctly identified. However, it focused more on ``format" than ``process'', the feature our manual labeling considered most important. Interestingly, Phi-3's second-highest focus on ``process'' may have contributed to its correct classification.

LIME added a qualitative layer by evaluating model interpretability, complementing quantitative metrics like precision, recall, and F1-score. For example, despite decent scores, RoBERTa often missed key features, while models like Phi-3 consistently aligned well with human annotations, showing stronger focus on relevant attributes. The focus on interpretability and performance emphasizes how important explainability tools like LIME are for testing machine learning models, especially in fields like NFR classification where accuracy and clarity are crucial.

\subsection{Data Analysis for Classifiers}
Table~\ref{tab:eval_comparison} compares the performance of various models in classifying non-functional requirements (NFRs). Among them, Gemma-2 achieved the best results, with a precision of 0.87, recall of 0.89, and F1-score of 0.88, reflecting its superior ability to detect true positives with minimal false positives. Phi-3 followed closely with a precision of 0.85, recall of 0.87, and F1-score of 0.86, making it a reliable and well-rounded classifier. Mistral-8B and Llama-3.1-8B also showed strong performance, with both models maintaining balanced metrics around the mid-80s in precision, recall, and F1-score, though slightly behind Gemma-2 and Phi-3. CodeBERT performed moderately, achieving a precision of 0.77, recall of 0.80, and F1-score of 0.78, indicating decent detection capabilities but with some misclassifications. In contrast, RoBERTa exhibited the weakest performance, with all metrics—precision, recall, and F1-score—falling around 0.30, highlighting its difficulty in understanding the task and correctly classifying NFRs. Overall, Gemma-2 and Phi-3 emerged as the most effective models, while RoBERTa lagged significantly behind.

\section{Discussion}
\label{appendix:discussion}


\subsection{Result Analysis}
\begin{figure*}[t]
    \centering
    \includegraphics[width=0.75\textwidth]{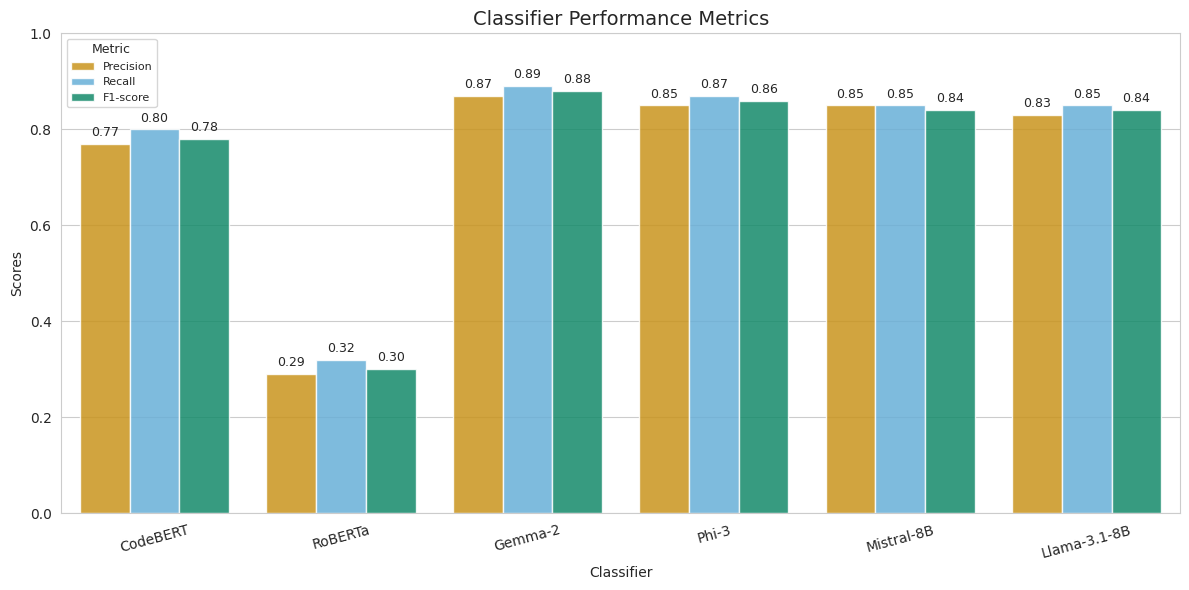}
    \caption{Comparison of Classifiers based on Precision, Recall, and F1-score}
    \label{result_analysis}
\end{figure*}

The comparative bar chart in Fig.~\ref{result_analysis} shows how well the classifier performed on three important metrics: F1-score, precision, and recall. The models' performance is balanced but unimpressive, as demonstrated by CodeBERT's consistent and fairly high scores throughout the measures. In contrast, RoBERTa is the least successful classifier in this investigation. This disparity demonstrates RoBERTa's internal shortcomings in managing the specific complexities of the evaluated task.

Gemma-2 emerges as the standout performer, achieving the highest Precision, Recall, and F1-score.Phi-3, that's close to it, has equally remarkable numbers that attest to its dependability as a top performer. Both Mistral-8B and Llama-3.1-8B demonstrate steady and reliable performance, with scores slightly trailing behind Gemma-2 and Phi-3. The results clearly underline Gemma-2’s dominance across evaluation metrics, reinforced by its seamless balance of precision and recall. Phi-3 follows as a close second. However, RoBERTa’s notably weak metrics remain a significant outlier, suggesting its inefficacy in this domain. When further analyzed using LIME Score Table~\ref{tab:lime_comparison}, additional insights into the models' interpretability and feature-detection accuracy emerge. Notably, Phi-3 secures the highest hit rate, demonstrating its exceptional ability to identify defining features within input data. Conversely, RoBERTa not only underperforms in primary metrics but also records the highest miss rate, further cementing its inadequacy in both performance and interpretability. Interestingly, while Gemma-2 shines in evaluation metrics, its hit rate in the LIME evaluation is slightly lower than Phi-3, indicating minor room for improvement in feature-level detection accuracy.

The LIME analysis affirms that Phi-3 and Gemma-2 consistently outperform others, showcasing high robustness in identifying key features. Meanwhile, the weaker performance of RoBERTa across all aspects both traditional metrics and interpretability clearly identifies it as unsuitable for this task.

\subsection{Techniques for Explainable AI (XAI)}

LIME analysis helps us understand how a model makes decisions by highlighting words or phrases in a requirement that influence its classification into a specific NFR category. For example, in the requirement “The system must handle 1,000 concurrent users,” LIME may identify “concurrent” and “1,000” as key terms affecting classification under performance. This allows us to check whether the model is focusing on relevant features or being misled by noise. 

Consider another example: “The system needs triple redundancy in terms of connection to the database.” Phi-3 assigns high importance scores to “redundancy” (0.76) and “connection” (0.69), while RoBERTa gives lower scores of 0.40 and 0.54, respectively. These scores reflect which features the models rely on for prediction. Higher scores from Phi-3 suggest it better captures fault tolerance concepts, while lower scores from other models—even if accurate—indicate difficulty in grasping sentence context.

\section{Threats to Validity}

Construct validity is threatened by the lack of developer input because we were unable to obtain information to validate classified NFRs. Furthermore, it can be difficult to identify requirements because project charters are frequently informal and unstructured.
In some cases, GitHub repositories lack detailed README sections, which can provide valuable insights into the project's goals and requirements, making it harder to extract relevant information. This limitation may pose a threat to internal validity. Additionally, bias in manual labeling could further affect internal validity, since expert opinions may influence term selection, thereby impacting LIME model assessments. Besides, external validity is threatened by the small sample of real-world project charters, even though we addressed this by adding new NFRs to the PROMISE\_exp dataset. This restricts our findings' applicability to larger projects and sectors. Furthermore, the robustness and accuracy of needs classification are impacted by the lack of thorough project charters.




\section{Conclusion}

Our research extended the NFR dataset using project charters and leveraged advanced LLMs - RoBERTa, CodeBERT, Gemma-2, Phi-3, Mistral-8B, and Llama-3.1-8B for classification. Metrics like precision, recall, F1-score, and LIME were used for evaluation, with Gemma-2 performing best, followed by Phi-3. Explainable AI ensured transparency and improved clarity in classification. However, some requirements overlap across categories, risking bias, and certain NFR types lacked sufficient data, affecting model reliability. To overcome these issues, we aim to expand the dataset, build a dedicated ML model, test newer LLMs like GPT-4 and Gemini, and continue using explainable AI for transparent results.

\bibliographystyle{plainnat}   

\bibliography{citation}


\vspace{12pt}

\end{document}